\documentclass[aip,apl,reprint]{revtex4-1}
 \usepackage{graphicx}
 \usepackage{amsmath}

\begin{document}

\title[Imaging vibrations]{Determining the vibrations between sensor and sample in SQUID microscopy}
\author{Daniel Schiessl}
\affiliation{Attocube systems AG,
K{\"o}niginstra{\ss}e 11a, 80539 Munich, Germany}

\author{John R. Kirtley}%
\affiliation{Geballe Laboratory for Advanced Materials, Stanford University, Stanford, California 94305-4045, USA}
\author{Lisa Paulius}
 \affiliation{Geballe Laboratory for Advanced Materials and Dept. of Applied Physics, Stanford University, Stanford, California 94305-4045, USA; Dept. of Physics, Western Michigan University, Kalamazoo, Michigan 49008-5252, USA}
\author{Aaron J. Rosenberg} 
\affiliation{Dept. of Applied Physics and Geballe Laboratory for Advanced Materials, Stanford University, Stanford, California 94305-4045, USA}
\author{Johanna C. Palmstrom} 
\affiliation{Dept. of Applied Physics and Geballe Laboratory for Advanced Materials, Stanford University, Stanford, California 94305-4045, USA}
\author{Rahim R. Ullah}
\affiliation{Dept. of Physics and Geballe Laboratory for Advanced Materials, Stanford University, Stanford, California 94305-4045, USA}
\author{Connor M. Holland} 
\affiliation{Dept. of Physics and Geballe Laboratory for Advanced Materials, Stanford University, Stanford, California 94305-4045, USA}
\author{Y.-K.-K. Fung}
\affiliation{IBM Research Division, T.J. Watson Research Center, Yorktown Heights, New York 10598, USA}
\author{Mark B. Ketchen}
\affiliation{OcteVue, Hadley, Massachusetts 01035, USA}
\author{Gerald W. Gibson, Jr.}
\affiliation{IBM Research Division, T.J. Watson Research Center, Yorktown Heights, New York 10598, USA}
\author{Kathryn A. Moler}
\affiliation{Dept. of Applied Physics, Dept. of Physics, and Geballe Laboratory for Advanced Materials, Stanford University, Stanford, California 94305-4045, USA}

\begin{abstract}
Vibrations can cause noise in scanning probe microscopies. Relative vibrations between the scanning sensor and the sample are important but can be more difficult to determine than absolute vibrations or vibrations relative to the laboratory. We measure the noise spectral density in a scanning SQUID microscope as a function of position near a localized source of magnetic field, and show that we can determine the spectra of all three components of the relative sensor-sample vibrations. This method is a powerful tool for diagnosing vibrational noise in scanning microscopies.
\end{abstract}

\maketitle

There is a large literature on detecting vibrational motion in scanning probe microscopy. Vibrations of the microscope as a whole have been determined using an acceleromater;\cite{quacquarelli2015spm} of the cantilever using piezoelectric sensing\cite{mahmoodi2008cft} or interferometry\cite{cretin1998mcv}; of the sample using stroboscopic optical microscopy,\cite{petitgrand2004smo} non-linear effects in atomic force microscopy,\cite{kolosov1993nld} or the cantilever deflection in scanning force microscopy.\cite{garcia2007mdc} In addition, a standard technique for analyzing resolution and stability in electron beam lithography is to move an anisotropically etched silicon edge relative to the beam.\cite{kratschmer1988qar} However there has been relatively less work on using sensor-sample vibrations as a diagnostic tool of vibrations within the microscope itself.

In this paper we show how one can determine all three components of the vibrations between sensor and sample by measuring the time dependence of the flux through a scanning Superconducting QUantum Interference Device (SQUID) pickup loop due to a superconducting vortex.

Our measurements were made in a scanning  microscope developed in a collaboration between Attocube and Stanford. Briefly, in this system the vibration isolation is provided by suspending the entire system from springs. The microscope is housed in a vacuum can that is inserted into a  liquid helium dewar; cooling is provided by He$_4$ exchange gas. The coarse positioning and scanning of the sample are performed by an Attocube piezoelectric stack. The SQUID is mounted on a cantilever consisting of a 3 mm wide, 10 mm long, and 25 $\mu$m thick copper shim, with wire bonds making electrical contacts. The SQUID mount and Attocube stack are mounted in a massive titanium housing. The titanium housing is suspended from a copper support which is firmly clamped to the sides of the vacuum can for thermalization of the microscope wiring. All measurements reported here were made at 4.2K. 

\begin{figure}[ht!]
\centering
\includegraphics[width=0.5\textwidth]{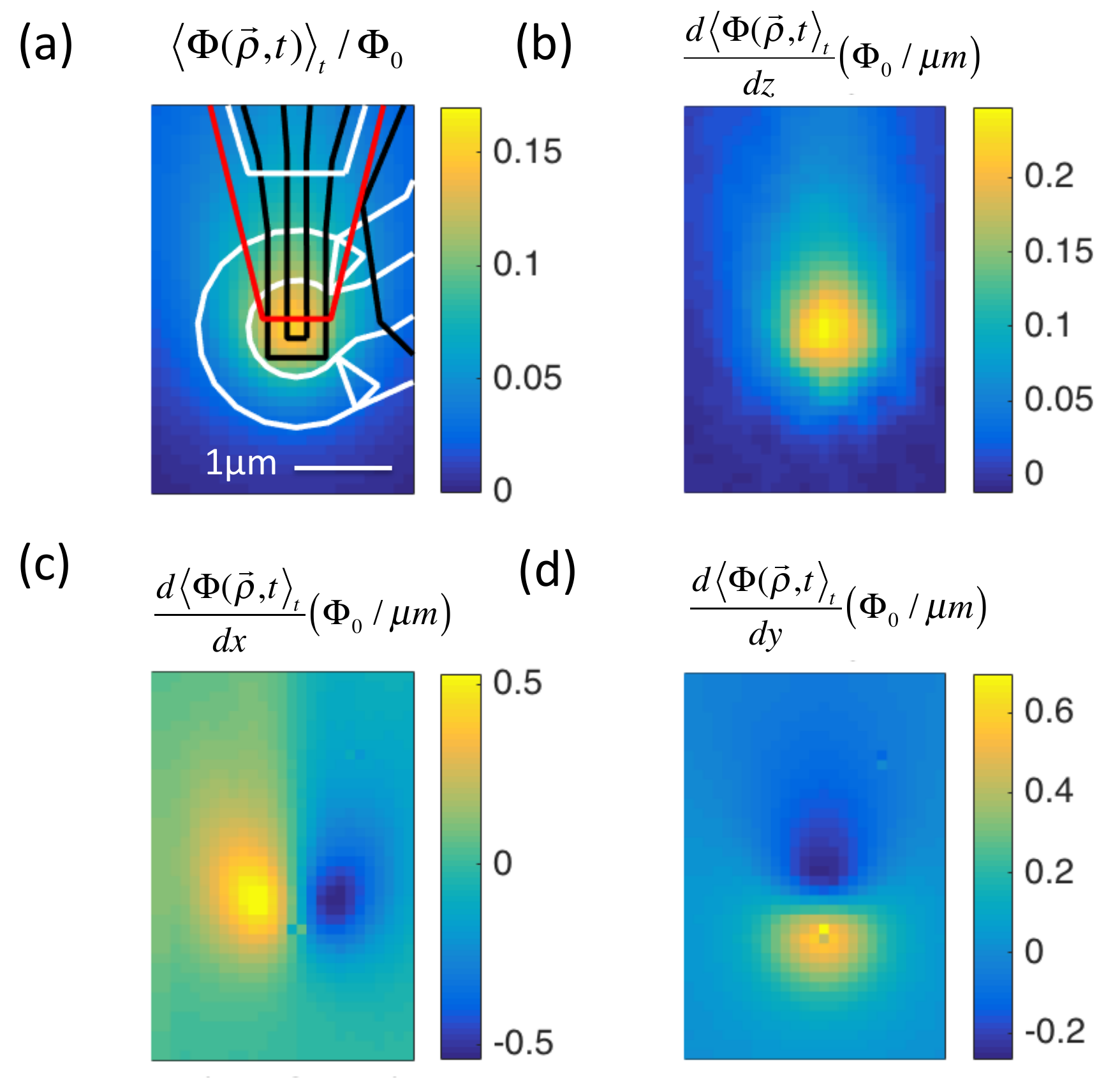}
\caption{Scanning SQUID microscopy of a superconducting vortex. (a) Spatially dependent, time averaged signal $\langle\Phi(\vec{\rho},t)\rangle_t/\Phi_0$ of a vortex trapped in a 0.4$\mu$m thick Nb film. Superimposed on the image is the layout of the pickup loop/field coil region of the 0.2$\mu$m inside pickup loop diameter SQUID susceptometer used. (b-d) The derivatives of the vortex flux image signal with respect to $z$, $x$, and $y$ respectively. }
\label{fig:vortex_derivatives}
\end{figure}

The SQUID susceptometer\cite{kirtley2016sss} used for these measurements has an integrated pickup loop and one turn field coil in the geometry indicated by the solid lines in Fig. \ref{fig:vortex_derivatives}a. We scan the sample relative to the SQUID's pick-up loop by applying a variable DC-Voltage to our piezo-based scanners. In our coordinate system the long axis of the cantilever is in the $\hat{y}$ direction, as are the leads to the pickup loop in the SQUID, and the sample plane is the $xy$ plane.

The studied sample is a 0.4 $\mu$m thick superconducting niobium film (T$_c$ = 9.2K). An isolated superconducting vortex was located by repeatedly cooling the sample in various externally applied fields until a low vortex density was achieved. This vortex acted as a field source for studying the vibrations in the system.

Magnetometry data was taken with the SQUID in a flux-locked loop,  at a data rate of 10 kHz, as 1-second time traces at each of 41x28x(0.08$\mu$m)$^2$ pixels. All flux noise and vibrational amplitudes in this paper are normalized in a 1 Hz bandwidth. 

Figure \ref{fig:vortex_derivatives}a displays $\langle \Phi(\vec{\rho},t) \rangle _t/\Phi_0$, the magnetic flux at scan position $\vec{\rho}=x\hat{x}+y\hat{y}$, averaged over the full time interval for each pixel, taken with a sensor to sample spacing of approximately 0.1 $\mu$m, and divided by the superconducting flux quantum $\Phi_0=h/2e$. Figures \ref{fig:vortex_derivatives}(b)-(d) are the derivatives of the time averaged flux signals from the vortex in the $\hat{z}$, $\hat{x}$, and $\hat{y}$ directions respectively. $d \langle \Phi(\vec{\rho},t) \rangle _t/dz$ was determined by retracting the SQUID 0.08 $\mu$m and subtracting two images; the other derivatives were taken numerically.

\begin{figure}[ht!]

\centering
\includegraphics[width=0.5\textwidth]{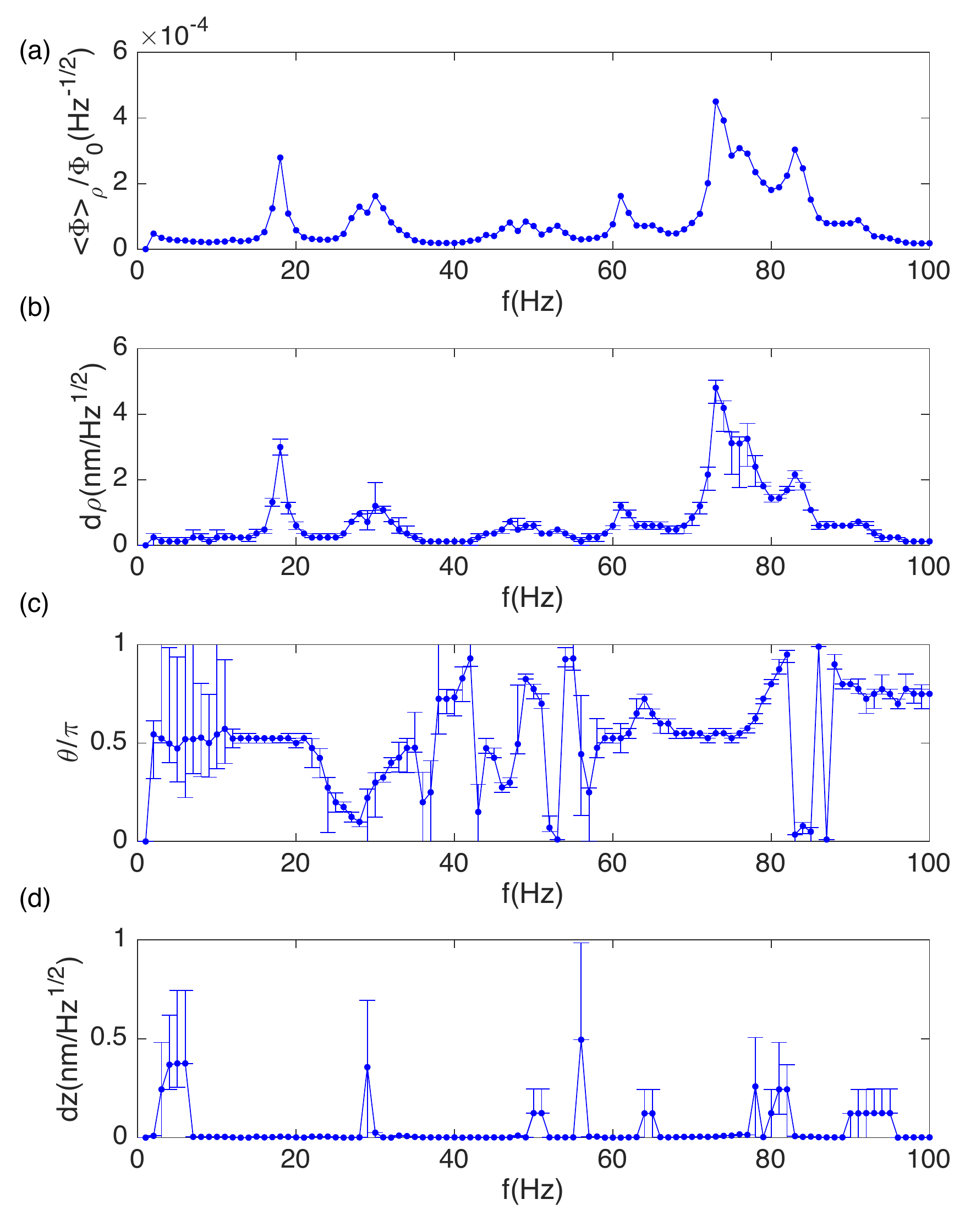}
\caption{Frequency dependence of the SQUID-sample vibrations. a) Average over the image of Fig. \ref{fig:vortex_derivatives}(a) of the Fourier components of the flux $|\Phi(f)|$, divided by the superconducting flux quantum $\Phi_0=h/2e$. (b-d) Fits of the components of the SQUID-sample vibrations in cylindrical coordinates.} 
\label{fig:fits_vs_f}
\end{figure}

\begin{figure}[ht!]
\centering
\includegraphics[width=0.5\textwidth]{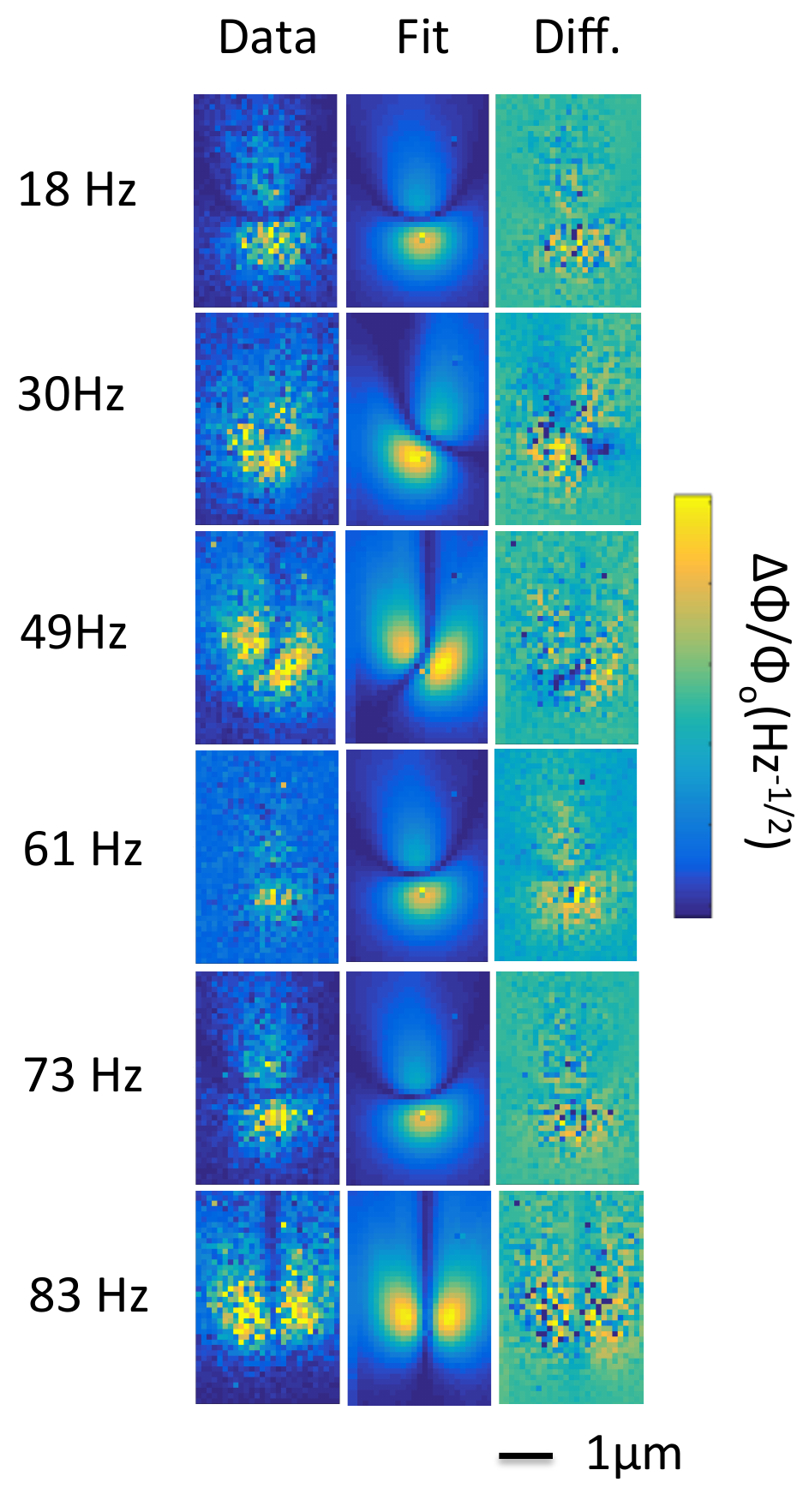}
\caption{Determination of sensor-sample vibrational vectors for 6 frequencies. The colormaps represent the power spectral densities $|\Phi(\vec{\rho},f)|$, at the frequencies labeled on the left, of the flux through the SQUID as a function of position relative to the center of the superconducting vortex. The data are in the left column, fits in the center, and the difference between the data and fits in the right column. The fit parameters and colorbar scales $\Delta\Phi/\Phi_0$ for these 6 modes are in Table \ref{table:fit_parameters}. The colorbar scales run from 0 to $\Delta\Phi/\Phi_0$ for the "Data" and "Fit" columns, but from $-\Delta\Phi/2\Phi_0$ to $\Delta\Phi/2\Phi_0$ for the "Diff." column.}
\label{fig:three_columns}
\end{figure}

Figure \ref{fig:fits_vs_f}a shows $\langle | \Phi(\vec{\rho},f) | \rangle_\rho$, the average over the entire image of Fig. \ref{fig:vortex_derivatives}a of the noise power density obtained by taking the absolute values of the discrete Fourier transforms of the time-traces for each pixel:
\begin{equation}
|\Phi(\vec{\rho},f)|=\frac{2T^{1/2}}{N} \left | \sum_{j=1}^{N}\Phi(\vec{\rho},t) \exp(-2\pi i (j-1)(k-1)/N) \right |,
\label{eq:dft}
\end{equation}
with the frequency $f=(k-1)/(2dt(N/2-1))$, $k=1:N/2$, the time $t=(j-1)dt/(N-1)$, $j=1:N$, the number of time samples $N=10000$, the total sampling time per pixel $T$= 1 s, and the time interval $dt=10^{-4}s$. 
 The flux noise amplitudes were much smaller when the SQUID sensor was moved away from the vortex. The flux noise had a striking spatial dependence that changed with frequency. Some examples are shown in Fig. \ref{fig:three_columns}.

We fit this data to a simple model, in which we expand the time dependence of the flux $\Phi(\vec{\rho},t)$ at each position $\vec{\rho}$ in the data set in a Taylor series around the equilibrium position of the vortex  $\vec{r_0}=x_0\hat{x}+y_0\hat{y}+z_0\hat{z}$. 
\begin{align}
\Phi(\vec{\rho},t) = \Phi(\vec{r}_0)+&\frac{d\langle\Phi(\vec{\rho},t)\rangle_t}{dx}(x(t)-x_0)+ \nonumber \\ \frac{d\langle\Phi(\vec{\rho},t)\rangle_t}{dy}(y(t)-y_0)+  
&\frac{d\langle\Phi(\vec{\rho},t)\rangle_t}{dz}(z(t)-z_0)+ ...,
\label{eq:taylor}
\end{align}
Fourier transforming the terms displayed in Eq. \ref{eq:taylor} results in
\begin{align}
|\Phi(\vec{\rho},f)| = & | \frac{d\langle\Phi(\vec{\rho},t)\rangle_t}{dx}dx(f)+\frac{d\langle\Phi(\vec{\rho},t)\rangle_t}{dy}dy(f)+ \nonumber \\
& \frac{d\langle\Phi(\vec{\rho},t)\rangle_t}{dz}dz(f) |,
\label{eq:taylorf}
\end{align}
where 
$dx(f)$, $dy(f)$, and $dz(f)$ are the noise powers of the sensor-sample motion for each frequency $f$. It is then possible to fit the measured Fourier components of the noise images at each frequency using the Cartesian parameters $dx(f),dy(f)$, and $dz(f)$ as fitting parameters. However, such a fitting procedure produces ambiguous results, with distinct volumes in the 3-dimensional parameter space with equivalently good chi-squared values $\chi^2(f) = \sum_{i}(\Phi_{\rm exp}(\vec{\rho_i},f) - \Phi_{\rm fit}(\vec{\rho_i},f))^2$. This ambiguity can be eliminated by casting the vibrational motion into cylindrical coordinates $\vec{dr}(f) = d\rho(f) ( \cos \theta(f) \hat{x} + \sin \theta(f) \hat{y}) + dz(f) \hat{z}$. Since we are taking absolute values,   
$\theta=\pi$ is equivalent to $\theta = 0$.
In cylindrical coordinates the flux signal is given by
\begin{align}
|\Phi(\vec{\rho},f)| =   | d\rho(f)  [ \cos\theta(f) \frac{d\langle\Phi(\vec{\rho},t\rangle_t}{dx}+ \nonumber \\ \sin\theta(f) \frac{d\langle\Phi(\vec{\rho},t\rangle_t}{dy}  ] + 
 dz(f) \frac{d\langle\Phi(\vec{\rho},t\rangle_t}{dz} | 
\label{eq:dphi}
\end{align}
This model describes the data reasonably well despite not allowing for phase shifts between the components of motion.

The flux derivatives displayed in Fig. \ref{fig:vortex_derivatives}(b)-(d) were inserted into Eq. \ref{eq:dphi} to fit the data with $d\rho(f),\theta(f)$ and $dz(f)$ as the three fitting parameters, using a matrix least squares search. Some representative results are shown in Fig. \ref{fig:three_columns}, with the corresponding fit parameters in Table \ref{table:fit_parameters}. The noise data in Fig. \ref{fig:three_columns} is characterized by lobes separated by lines of nodes, which are well reproduced by the fits. There are some systematic differences between the data and the fits, which could perhaps be reduced by allowing two more fitting parameters for the relative phase shifts between the vibrational components.

Similar fit parameters for all frequencies in the 0-100 Hz range are displayed in Fig. \ref{fig:fits_vs_f}(b)-(d). Error bars were assigned by statistical bootstrapping. \cite{efron1986bmf} 
Briefly, in this analysis a random sampling of the data was generated, with substitutions, to produce the same number of points as the original set. This set was fit to the model allowing all three parameters to vary, best fit parameters were recorded, and the procedure was repeated 200 times. A histogram of the best fit parameters was generated, and confidence interval limits were set at the 2.5\% and 97.5\% levels. The best fit values for $\theta$ are dependent on frequency (see Fig. \ref{fig:fits_vs_f}c), with the largest vibrational amplitudes in-plane, with $\theta \approx \pi/2$: in the $\hat{y}$ direction. This is surprising, since the lowest vibrational frequency of the cantilever should be in the $\hat{z}$ direction. In addition, since the long axis of the cantilever is in the $\hat{y}$ direction, one might expect a higher resonance frequency and lower vibrational amplitude in that direction.

\begin{table}[ht]

\centering 
\begin{tabular}{| c | c | c | c | c |}

\hline                        
f (Hz) & $\Delta\Phi/\Phi_0(Hz^{-1/2})$ & d$\rho$(nm/Hz$^{1/2}$) &  $\theta$($^o$) & dz(nm/Hz$^{1/2}$)  \\

\hline  
18   & $4.5\times 10^{-3}$   & 3.0$\pm$0.3   & 95$\pm$2   & 0.1 $\pm$0.2   \\
30   & $1.5\times 10^{-3}$   & 1.3$\pm$0.4   & 62$\pm$20   & 0.1$\pm$1.2    \\
49   & $7\times 10^{-4}$     & 0.6$\pm$0.1   & 149$\pm$6  & 0.0 $\pm$0.6   \\
61   & $1.6\times 10^{-3}$   & 1.2$\pm$0.2   & 99$\pm$9   & 0.0 $\pm$0.2   \\
73   & $7\times 10^{-3}$     & 4.8$\pm$0.4   & 99$\pm$2   & 0.0 $\pm$0.2   \\
83   & $2.3\times 10^{-3}$   & 2.1$\pm$0.1   & 178$\pm$3  & 0.0 $\pm$ 0.4   \\
[1ex] 
\hline 
\end{tabular}
\caption{Parameters for Figure \ref{fig:three_columns}.}
\label{table:fit_parameters} 
\end{table}

We are able to determine all three components of the frequency dependent vibrational motion between sensor and sample with a precision of a few tenths of a nm using this technique. Although our measurements were made with a SQUID microscope imaging a superconducting vortex, the same technique could be used with any scanning probe sensing a localized source.

\section*{Acknowledgements}

This work was supported by an NSF IMR-MIP Grant No. DMR-0957616. J.C.P. was supported by a Gabilan Stanford Graduate Fellowship and an NSF Graduate Research Fellowship, Grant No. DGE-114747. R.R.U. and C.M.H. were supported by VPUE and the Departments of Physics and Applied Physics at Stanford University. We would like to thank Micah J. Stoutimore for providing the Nb film used to study vortices, and Zheng Cui for useful conversations.

\bibliography{./bibliography/pairing_symmetry}
\end{document}